\documentclass[onecolumn,amsmath,amssymb,aps]{revtex4}
\usepackage{bm}
\usepackage{graphicx}
\usepackage{dcolumn}
\usepackage{epsfig}
\usepackage[cp1251]{inputenc}
\usepackage[english]{babel}
\usepackage{euscript}
\usepackage{amstext}
\usepackage{amsmath}
\usepackage{amssymb}
\usepackage{subfigure}
\usepackage{wrapfig}
\DeclareGraphicsExtensions{.pdf,.jpg,.eps}

\begin{document}

\preprint{}

\title{Observation of diffraction with the CMS experiment at the Large Hadron Collider}

\thanks{\sl Presented at the Workshop on Forward Physics at the LHC (December 12-14, 2010), Manchester, United Kingdom}

\author{Dmytro Volyanskyy\footnote{on behalf of the CMS collaboration}}
 \email{Dmytro.Volyanskyy@cern.ch}
\affiliation{%
Deutsche Elektronen-Synchrotron DESY \\
Notkestrasse 85, 22607 Hamburg, Germany
}%

\begin{abstract}
A clear evidence of inclusive diffraction observed by the CMS detector at the Large Hadron Collider 
in minimum bias events at $\sqrt{s}=0.9$~TeV, $2.36$~TeV is presented. 
%
The observed diffractive signal is dominated by inclusive single-diffractive dissociation 
and can be identified by the presence of a Large Rapidity Gap that extends over the forward region of the CMS detector. 
A comparison of the data with Monte Carlo predictions provided by PYTHIA6 and PHOJET generators is given.
In addition, first observation of the single-diffractive production of di-jets at $\sqrt{s}=7$~TeV is presented. 
\begin{description}
\item[PACS numbers:] 
\item[Keywords:] CMS, diffraction
\end{description}
\end{abstract}
\maketitle

\section{Physics Motivation}
Diffractive scattering processes attract a lot of interest due to a significant contribution
to the total $pp$ cross-section~[1].
A proper constraint on the diffractive component is thus essential
to improve our understanding of collision data and the pile-up as well as to tune
the existing Monte Carlo models of $pp$ interactions at the LHC.
In $pp$ collisions, a diffractive process is mainly a reaction $pp \rightarrow X Y$,
where $X$ and $Y$ can either be protons or low-mass systems which may be
a resonance or a continuum state. In all cases, the final states $X$ and $Y$
acquire the energy approximately equal to that of the incoming protons and
carry the quantum numbers of the proton as well as are separated by a Large Rapidity Gap~(LRG).
Diffraction in the presence of a hard scale can be described with perturbative QCD
by the exchange of a colorless state of quarks or gluons, whereas soft diffraction
at high energies is described in the Regge Theory~[2] as a colorless exchange mediated
by a specific trajectory, the Pomeron, having the quantum numbers of the vacuum.
Two main types of diffractive processes occurring in $pp$ collisions
are the single-diffractive dissociation~(SD) where one of the protons dissociates
and the double-diffractive dissociation~(DD) where both protons are scattered into a low-mass system.
\begin{figure}[h!]
\begin{center}
\resizebox{6.9in}{!}{
\rotatebox{0}{
\includegraphics{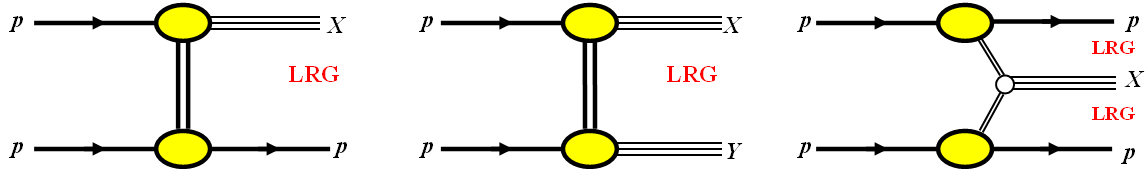}}}
\caption{ \sl Diagram of the SD reaction $pp \rightarrow p X$~(left), the DD reaction $pp \rightarrow X Y$~(middle) and the CD reaction $pp \rightarrow p X p$~(right).}
\end{center}
\end{figure}
\begin{table}[hb!]
\caption{\sl The estimated cross-sections for different $pp$ interactions at the LHC for $\sqrt{s}=14$~TeV{\rm~[1]}.}     
\begin{center}
\begin{tabular}{|c|c|}
\hline
Total $pp$ cross-section         &  $111.5\pm1.2^{+4.1}_{-2.1}$~mb \\  \hline \hline
Inelastic, non-diffractive interactions &  $\sim65$~mb \\  \hline \hline
Elastic scattering                      &  $\sim30$~mb \\  \hline \hline
SD dissociation ($pp \rightarrow p X$) &  $\sim10$~mb \\  \hline 
DD dissociation ($pp \rightarrow X Y$) &  $\sim7$~mb  \\  \hline 
CD dissociation ($pp \rightarrow p X p$) & $\sim1$~mb   \\  \hline 
\end{tabular}
\end{center}
\end{table}
Another diffractive process that may occur in $pp$ collisions with relatively large cross-section 
is the central-diffractive dissociation~(CD), also known as Double Pomeron Exchange, where the final state includes 
two incoming protons and one low-mass system created as a result of the interaction 
between two Pomeron-like objects emitted by the protons.
Figure~1 illustrates a sketch for each of these processes and Table~I gives 
the corresponding cross-sections at the LHC for $\sqrt{s}=14$~TeV.

It is known that the cross-section of hard diffractive processes can be factorized into a hard scattering contribution 
and a diffractive parton distribution function~(dPDF), which contains  
valuable information about low-$x$ partons. However, the factorization is broken
when scattering between spectator partons of the beam particles takes place. 
This effect is quantified by the so-called rapidity gap survival probability, 
which can be probed by measuring the ratio of diffractive to inclusive processes with the same hard scale.
At the Tevatron, the ratio is found to be $O(1\%)$, whereas theoretical expectations
at the LHC vary from a fraction of a percent to up to $30\%$~[3].
A good way to measure this quantity at the LHC is provided by the SD production of $W$ and di-jets, 
which are hard diffractive processes sensitive to the quark and gluon component of the proton dPDF, correspondingly. 
A selection of such events at CMS can be performed using the multiplicity distributions of calorimeter towers or tracks 
in the central region of CMS and calorimeter towers in the forward CMS calorimeters exploiting the fact that diffractive events
on average have lower multiplicity in the central region and in the $"$gap side$"$ than non-difractive ones.
Feasibility studies to detect the SD productions of $W$~[4] and di-jets~[5] at CMS
showed that the diffractive events peak in the regions of no activity 
in the forward calorimeters, which is due to the LRG presence.

\section{Experimental instrumentation}

The Compact Muon Solenoid~(CMS) experiment built at the Large Hadron Collider 
is one of the largest scientific instruments ever constructed. 
The detector has a cylindrical structure with the overall diameter of $15$~m, 
the overall length of $21$~m and the total weight of about $12.5$~thousands tons 
and consists of about $76.5$ millions of readout channels in total.
Its detailed description can be found in [6].
The detector has been designed, constructed and currently operated by the collaboration consisting of more 
than $3500$ scientists from $38$ countries.
To enhance the physics reach of the experiment the CMS detector includes 
several calorimeters covering the very forward region of the experiment. 
This significantly expands the CMS capability to investigate physics processes 
occurring at very low polar angles and so, provides a valuable tool to study 
diffractive scattering, low-$x$ QCD, multi-parton interactions and underlying event structure. 
\begin{figure}[t!]
  \begin{center}
     \resizebox{0.85\textwidth}{!}{
    \includegraphics[width=0.68\textwidth]{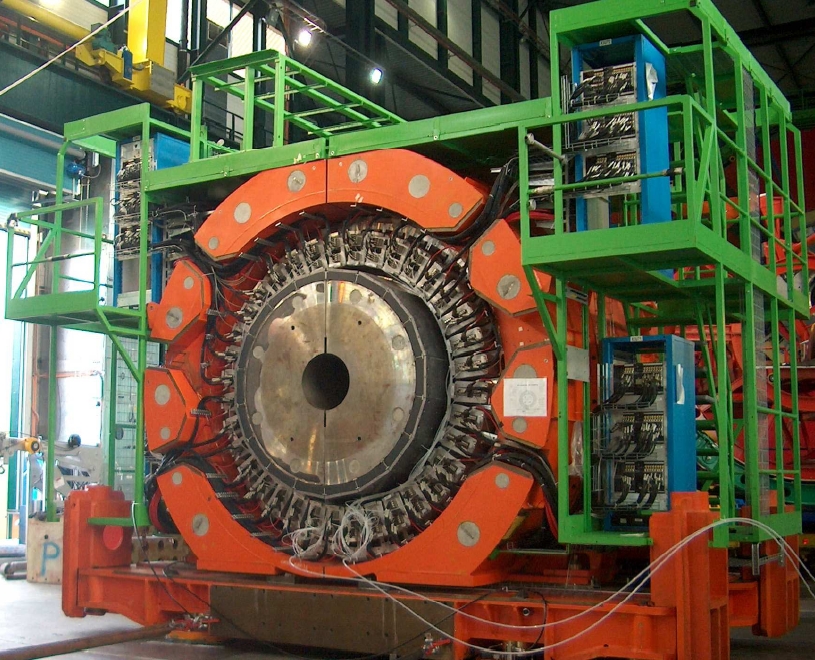}
    \includegraphics[width=0.72\textwidth]{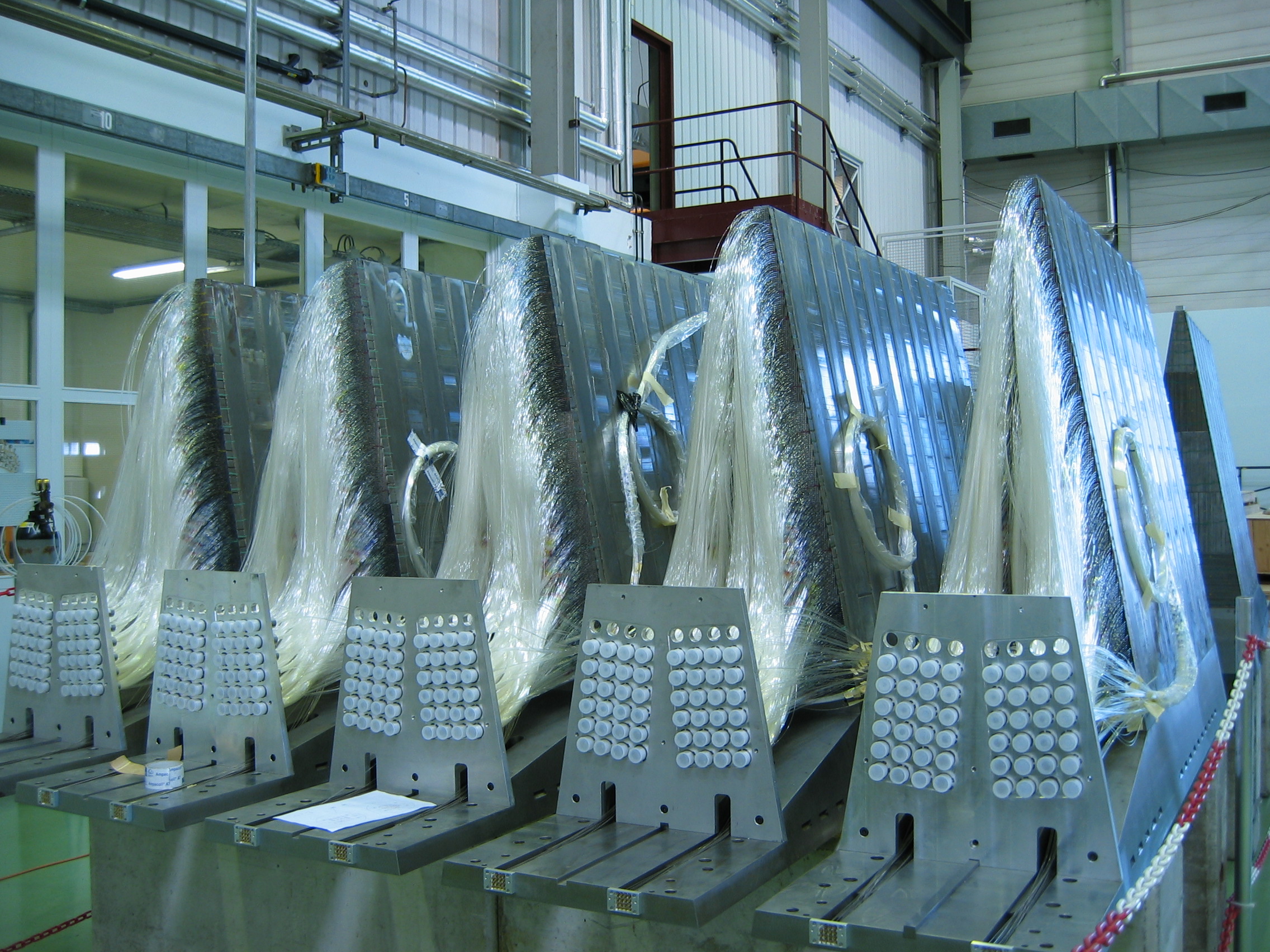} 
   }
   \end{center}
  \caption{\sl One of the two HF calorimeters before the installation into the CMS interaction point IP5 ~(left) and HF wedges during assembly~(right).}
\end{figure}

One of the forward detectors, which was used in particular to observe  
diffractive scattering in first minimum bias events, is the Hadronic Forward~(HF) detector~[7] shown in Figure~2. 
It includes two calorimeters HF+ and HF-- which are located at a distance of $11.2$~m 
on both sides from the IP5 and covering the pseudorapidity range $3<|\eta|<5$. 
The HF is a Cerenkov sampling calorimeter which uses radiation hard
quartz fibers as the active material and steel plates as the absorber.
It is azimuthally subdivided into $20^{0}$ modular wedges, 
each of which consists of two azimuthal sectors of $10^{0}$.
The detector fibers run parallel to the beamline and 
are bundled to form $0.175 \times 1.175$ ($\Delta\eta \times \Delta\phi$) towers.
Half of the fibers run over the full depth of the absorber,
whereas the other half starts at a depth of $22$~cm from the front of the
detector. These two sets of fibers are read out separately.
Such a structure allows to distinguish showers
generated by electrons and photons, which deposit a large
fraction of their energy in the first $22$~cm, from those generated
by hadrons, which produce signals in both calorimeter
segments.  The detector extends over $10$ interaction lengths and includes
$1200$ towers in total. Its main physics objective is to measure 
the forward energy flow and forward jets.

The other CMS subcomponents that have been used to observe 
diffractive scattering in first minimum bias events 
are the ECAL and HCAL calorimeters covering the central region of CMS $-3<\eta<3$. 
These are the barrel and endcap detectors, which provide high-precision
measurements of the energy of collision-products. 
The ECAL has an energy resolution of better than $0.5\%$ above $100$~GeV~[8], 
while the HCAL, when combined with the ECAL, is able to measure 
the energy of hadrons with resolution of better than $10\%$ above $300$~GeV~[9].
To trigger the CMS readout, two elements of the CMS detector monitoring system -- 
the Beam Scintillator Counters~(BSC) and the Beam Pick-up Timing for the eXperiments~(BPTX) 
were used. The BSC devices are located at a distance of $10.86$~m on both sides 
from the interaction point covering the pseudorapidity range 
$3.23<|\eta|<4.65$ and providing hit and coincidence signals
with a time resolution of about $3$~ns. Each BSC comprises $16$ scintillator tiles. 
Two BPTX elements are located around the beam pipe at a distance of $\pm175$~m from
the interaction point providing precise information on the bunch structure and timing
of the incoming beam with a time resolution better than $0.2$~ns.

\section{Event selection and acceptance}

A search for diffractive events has been made as soon as the CMS detector has started to take collision data.
In this paper, early observation of inclusive diffraction in minimum bias events 
collected by the CMS detector at $\sqrt{s}=0.9$~TeV and $2.36$~TeV  in the fall of 2009~[10] is presented.
To select a sample with the largest possible acceptance for SD events while suppressing beam-related background 
the following conditions were imposed. First, the presence of both beams was required 
by requesting the coincidence between BPTX signals in conjunction with a signal in one of the BSCs 
(the coincidence of the BSC signals was not required, since it would have obviously suppressed SD events). 
Then, to ensure that the selected event is a collision candidate, the events were required to have at least one primary vertex reconstructed 
from at least $3$ tracks with a $z$ distance to the interaction point below $15$~cm and 
a transverse distance from the $z$-axis smaller than $2$~cm. Further cuts were applied to reject beam-scrapping and beam-halo events 
as well as events with large signals consistent with noise in the HCAL.
Finally, the energy threshold in the calorimeters was set to $3$~GeV, except for the HF where the threshold of $4$~GeV was used.
After applying all the requirements the number of selected events are 207345 and 11848 at $\sqrt{s}=0.9$~TeV and $2.36$~TeV, correspondingly.

To understand the effect of the applied selection cuts, the acceptance for SD events 
was studied as a function of the generated value of $\xi$ -- the fractional energy loss of the scattered proton, 
which in turn is the fraction of the incoming proton energy carried by the Pomeron.
The corresponding distributions obtained with PYTHIA6 and PHOJET 1.12-35 generators 
for both collision energies are shown in Figure~3. The figure also illustrates 
the generator-level $\xi$ distributions of SD events, which peak at low $\xi$ values 
following roughly the $1/\xi$ dependence.
\begin{figure}[b!]
  \begin{center}
     \resizebox{1.\textwidth}{!}{
       \rotatebox{90}{
    \includegraphics[width=1.\textwidth]{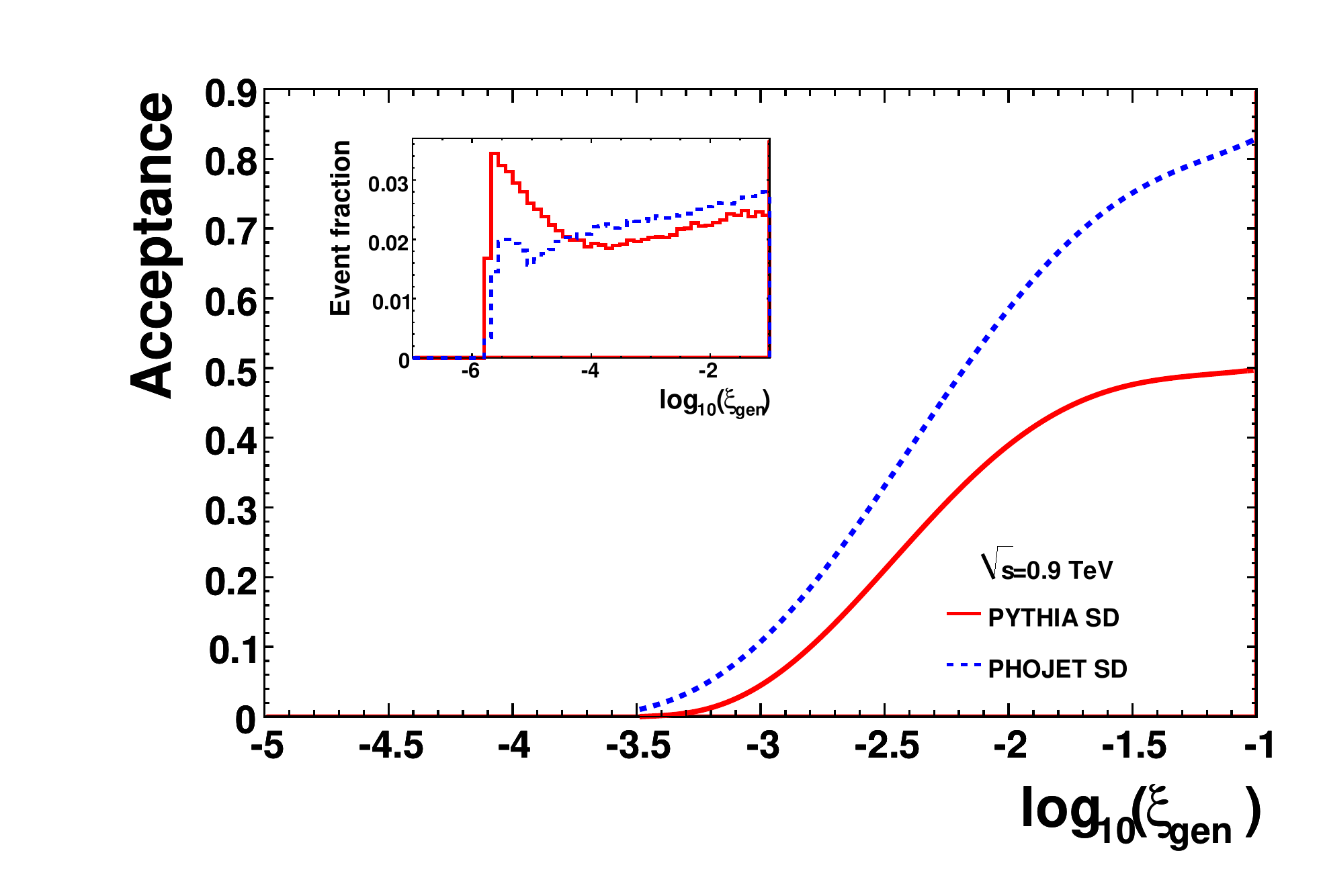}}
       \rotatebox{90}{
    \includegraphics[width=1.\textwidth]{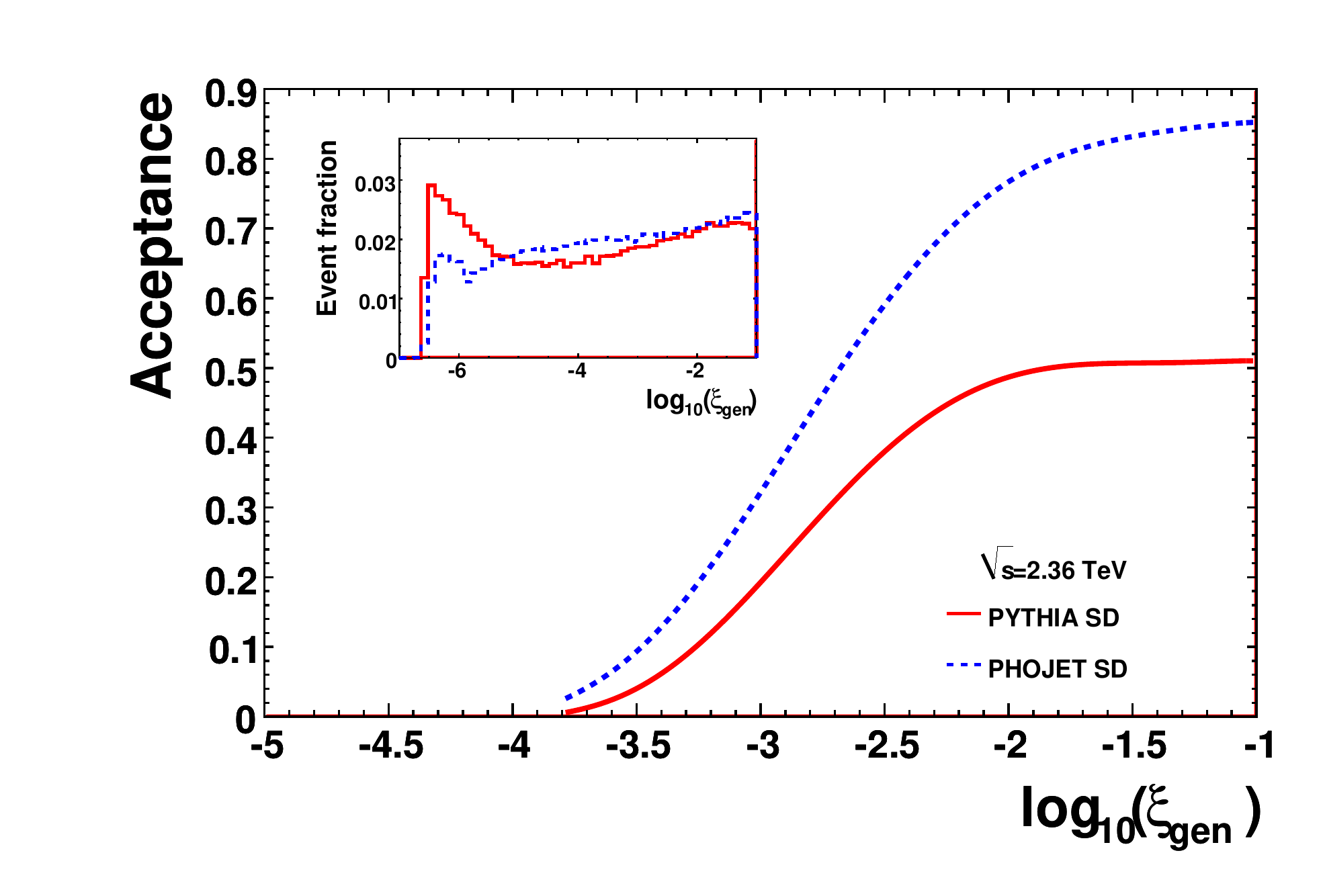}} 
   }
   \end{center}
  \caption{\sl Acceptance for SD events obtained with PYTHIA and PHOJET after applying the selection cuts at $\sqrt{s}=0.9$~TeV~(left) and $2.36$~TeV~(right) 
              as a function of the generated value of $\xi$. The generator-level $\xi$ distributions are illustrated in the insert. The discrepancy between the generators is               described in text.}
\end{figure}
As can be seen, there is a large discrepancy between the PYTHIA and PHOJET distributions
which can be explained by the fact that the two generators model the diffractive contribution in a different way.
In particular, PHOJET uses the Dual Parton Model~[11] which describes diffractive processes by the interaction
of the Pomeron with hadrons or another Pomeron. In this generator, the Pomeron exchange is represented as a mixture
of soft and hard contributions, where the latter are described by perturbative QCD matrix elements.
In contrast, PYTHIA6  does not include hard diffractive contributions and 
uses the Schuler-Sj\"{o}strand model to describe the diffractive cross-sections and event characteristics~[12].
Another explicit observation from the obtained distributions is a small acceptance for low-$\xi$ events at both energies. 
To explain this feature one needs to take into account the fact that
$\xi \sim M_{X}^2$, where $M_{X}$ is the total mass of the diffractive system $X$. 
Obviously, the lighter the system $X$ the harder to detect it experimentally.
The selection efficiency for SD events is found to be about $35\%$ according to PHOJET and about $20\%$ according to PYTHIA, 
whereas for non-diffractive events it is approximately equal to $85\%$ for both generators. 

\section{Observation of inclusive diffraction}
Due to the presence of a LRG in the final state, SD events can be identified
as those having no (or low) activity on one side of the forward region of CMS. 
As a result, the diffractive signal can be observed when plotting the selected events 
as a function of the energy deposition and tower multiplicity in the HF. 
Another variable giving an obvious evidence of the diffractive contribution is $E \pm p_{\rm z} = \sum_{i} (E_{i} \pm p_{z,i})$, 
where $E_{i}$ is the tower energy, $p_{z,i}$ is the longitudinal momentum and the sum runs over all calorimeter 
towers reconstructed in an event. The plus~(minus) sign is assigned 
when the proton emitting the Pomeron moves in the $+z$~($-z$) direction. 
Using energy and longitudinal momentum conservation it can be shown that 
this variable is roughly equal twice the Pomeron energy and therefore, is directly proportional to $\xi$.  
Taking into account the fact that the diffractive cross-section peaks at small $\xi$, diffractive events should cluster at low values of $E \pm p_{\rm z}$.
Figures 4 and 5 illustrate the distributions of the selected events as a function of $E + p_{\rm z}$  and the energy deposition in the HF+, $E_{HF+}$, 
at $\sqrt{s}=0.9$~TeV and $2.36$~TeV, respectively. The other variable, the multiplicity of towers above threshold in the HF is shown in Figure~6 for both energies. 
\begin{figure}[t!]
  \begin{center}
     \resizebox{0.95\textwidth}{!}{
    \includegraphics[width=1.0\textwidth]{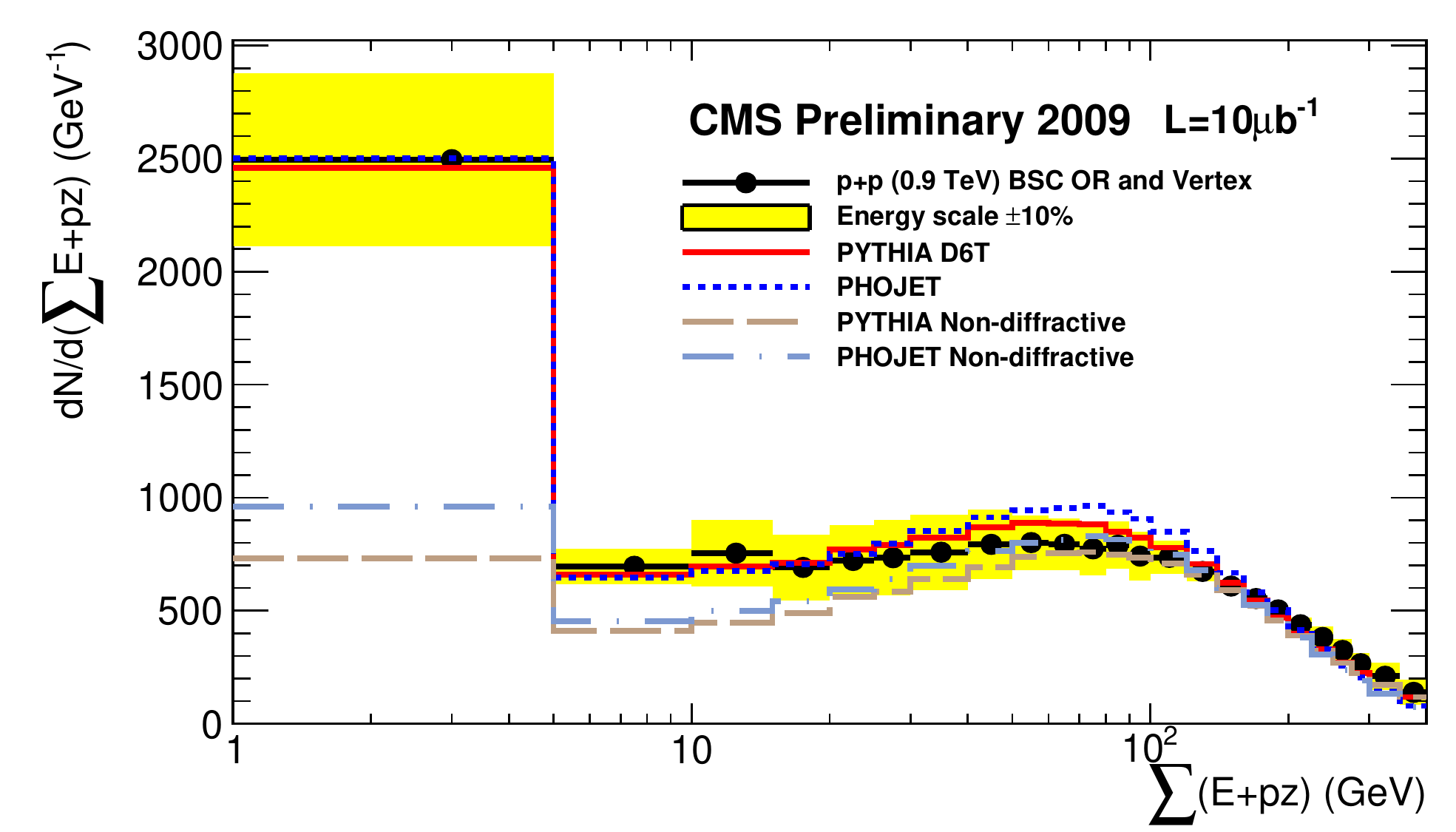}
    \includegraphics[width=1.0\textwidth]{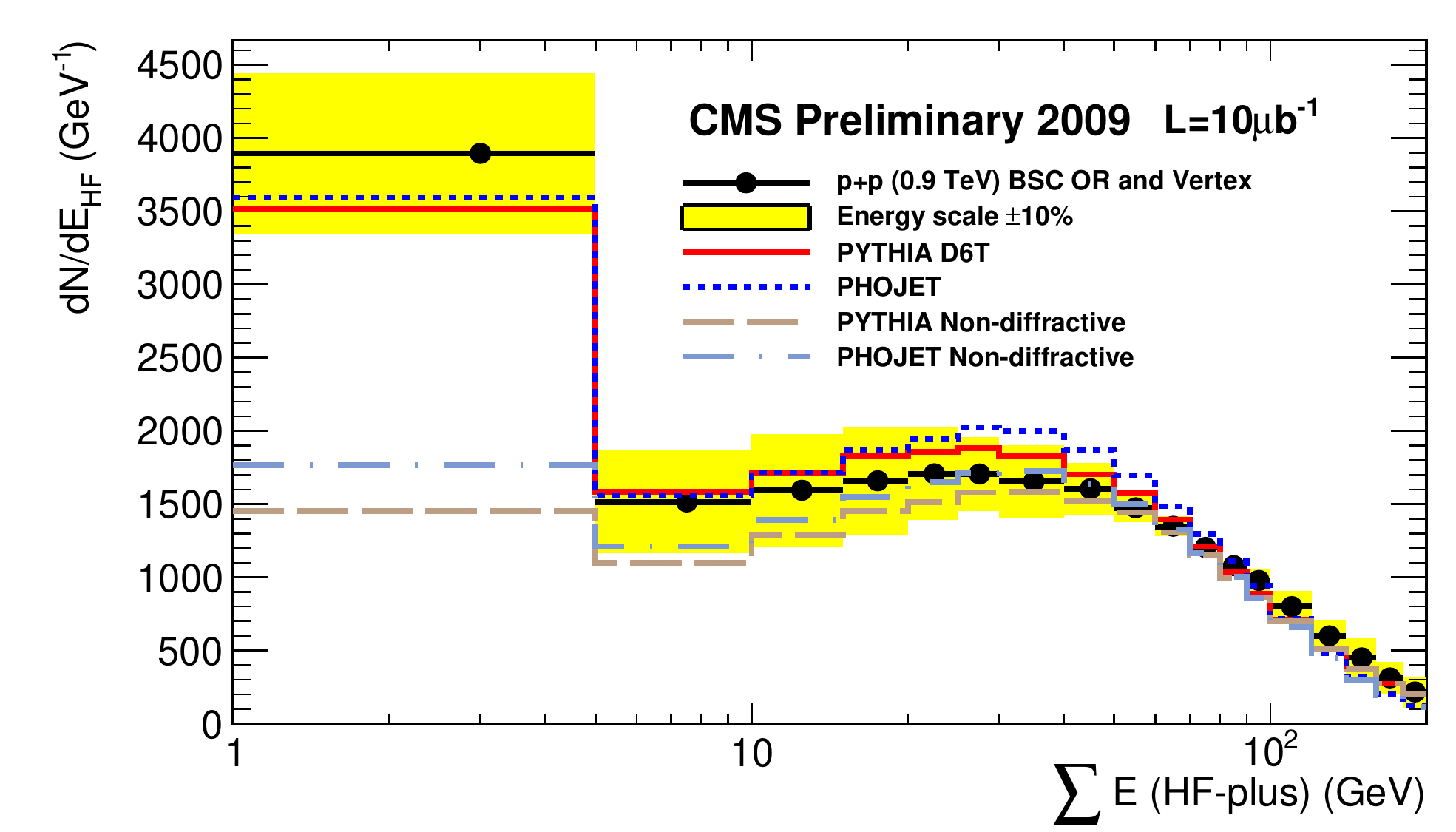}
   }
   \end{center}
  \caption{\sl Distributions of the uncorrected variables $E+p_{\rm z}$~(right) and $E_{HF+}$~(left) at $\sqrt{s}=0.9$~TeV. See text for details.}
\end{figure}
\begin{figure}[h!]
  \begin{center}
     \resizebox{0.95\textwidth}{!}{
    \includegraphics[width=1.0\textwidth]{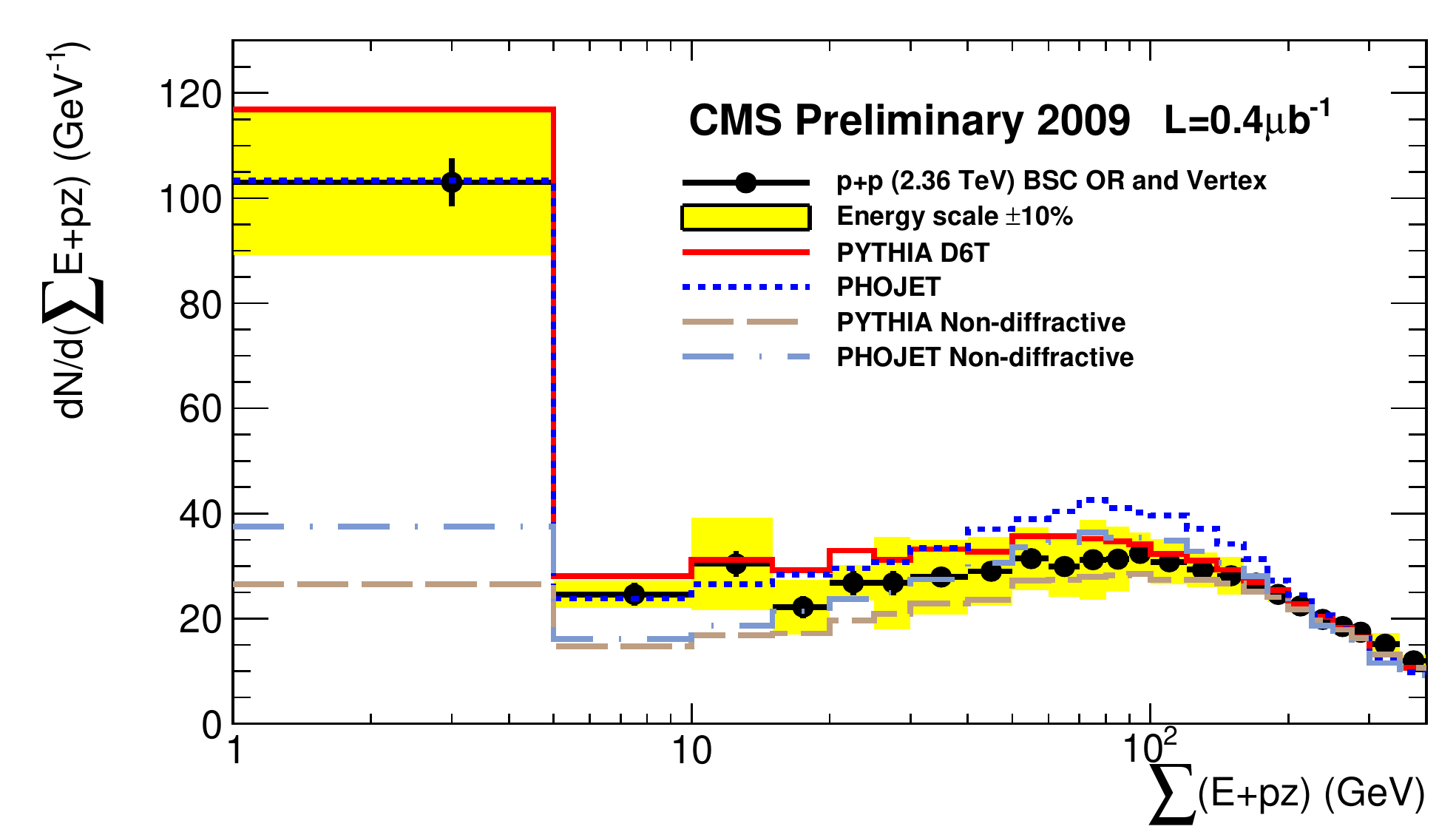}
    \includegraphics[width=1.0\textwidth]{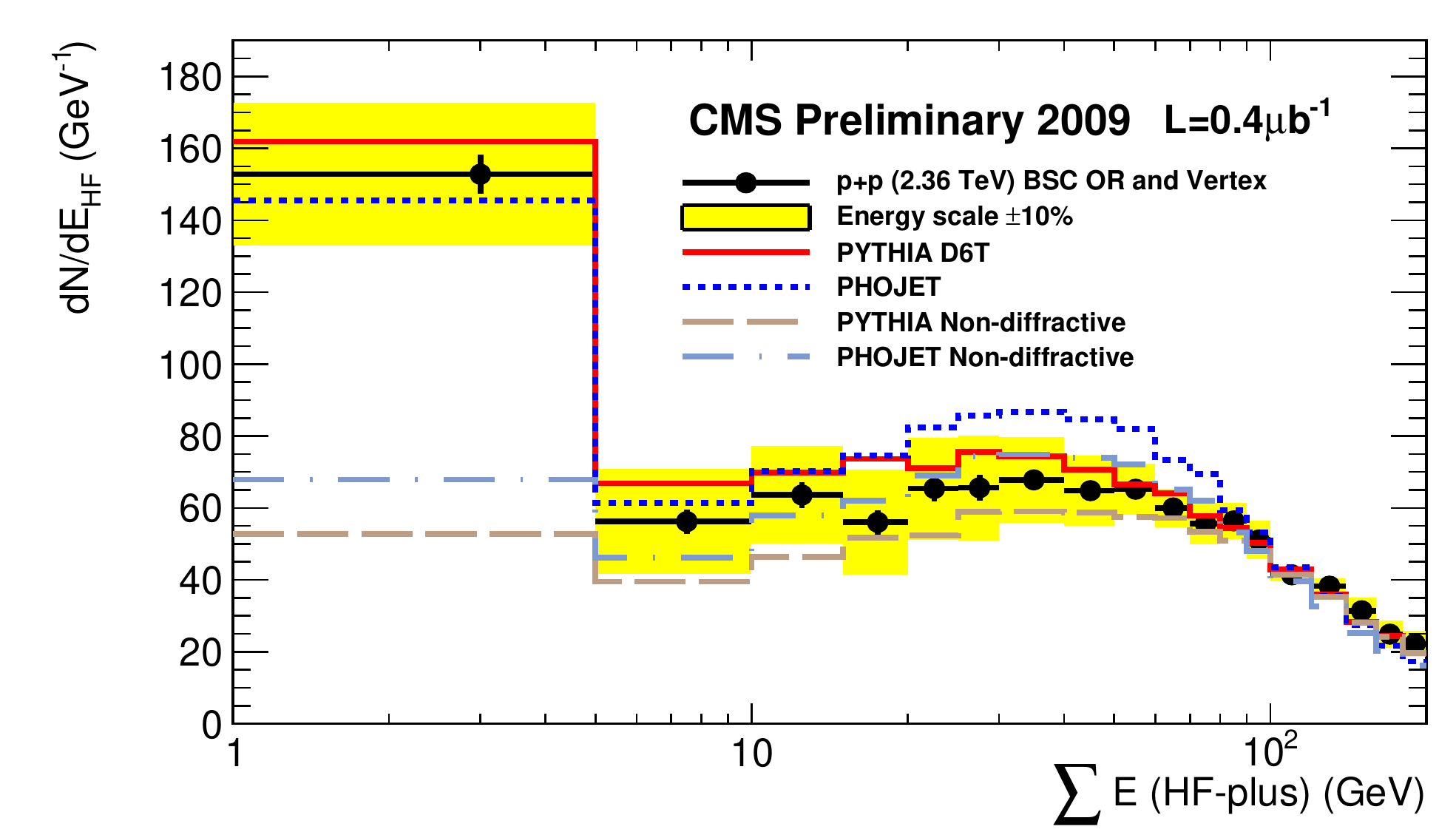}
   }
   \end{center}
  \caption{\sl Distributions of the uncorrected variables $E+p_{\rm z}$~(right) and $E_{HF+}$~(left) at $\sqrt{s}=2.36$~TeV. See text for details.}
\end{figure}  
%
%
As can clearly be seen, the diffractive signal peaks mainly in the first bin of the presented distributions. 
The vertical bars illustrate the statistical uncertainty of the data, whereas 
the bands demonstrate the main systematic uncertainty, which is due to the current imperfect calibration 
of the detectors and is estimated by a $10\%$ variation of the energy scale. The corresponding PYTHIA and PHOJET predictions normalized to the data
are also shown in these figures. As can be observed, there is a reasonable agreement between the data and Monte Carlo simulated events. 
In overall, PYTHIA describes the non-diffractive part of the spectra better than PHOJET. 
The predictions given by the generators without diffractive contribution are demonstrated in addition to strengthen the evidence of diffractive dissociation.
\begin{figure}[t!]
  \begin{center}
     \resizebox{0.95\textwidth}{!}{
    \includegraphics[width=1.0\textwidth]{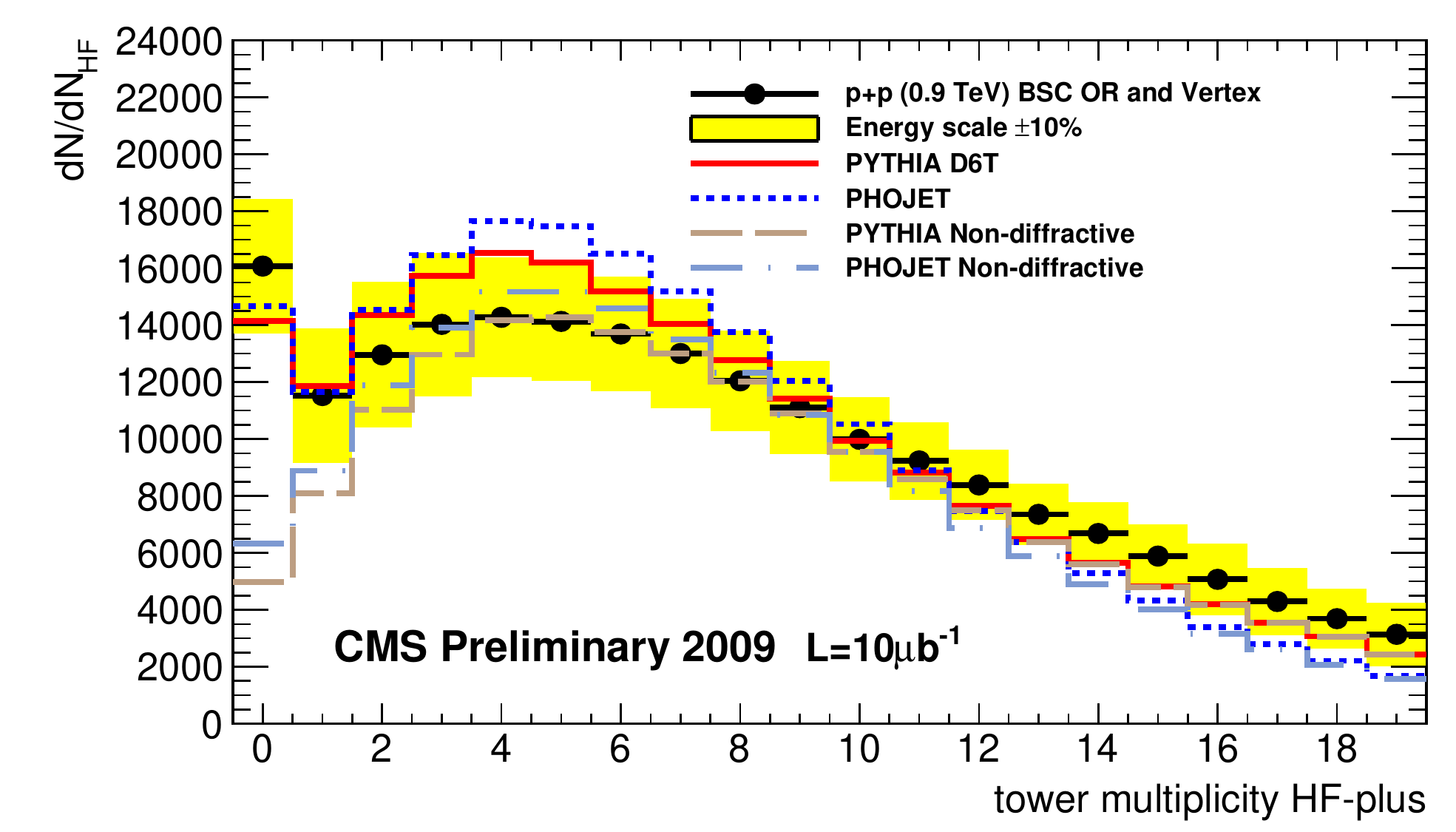}
    \includegraphics[width=1.0\textwidth]{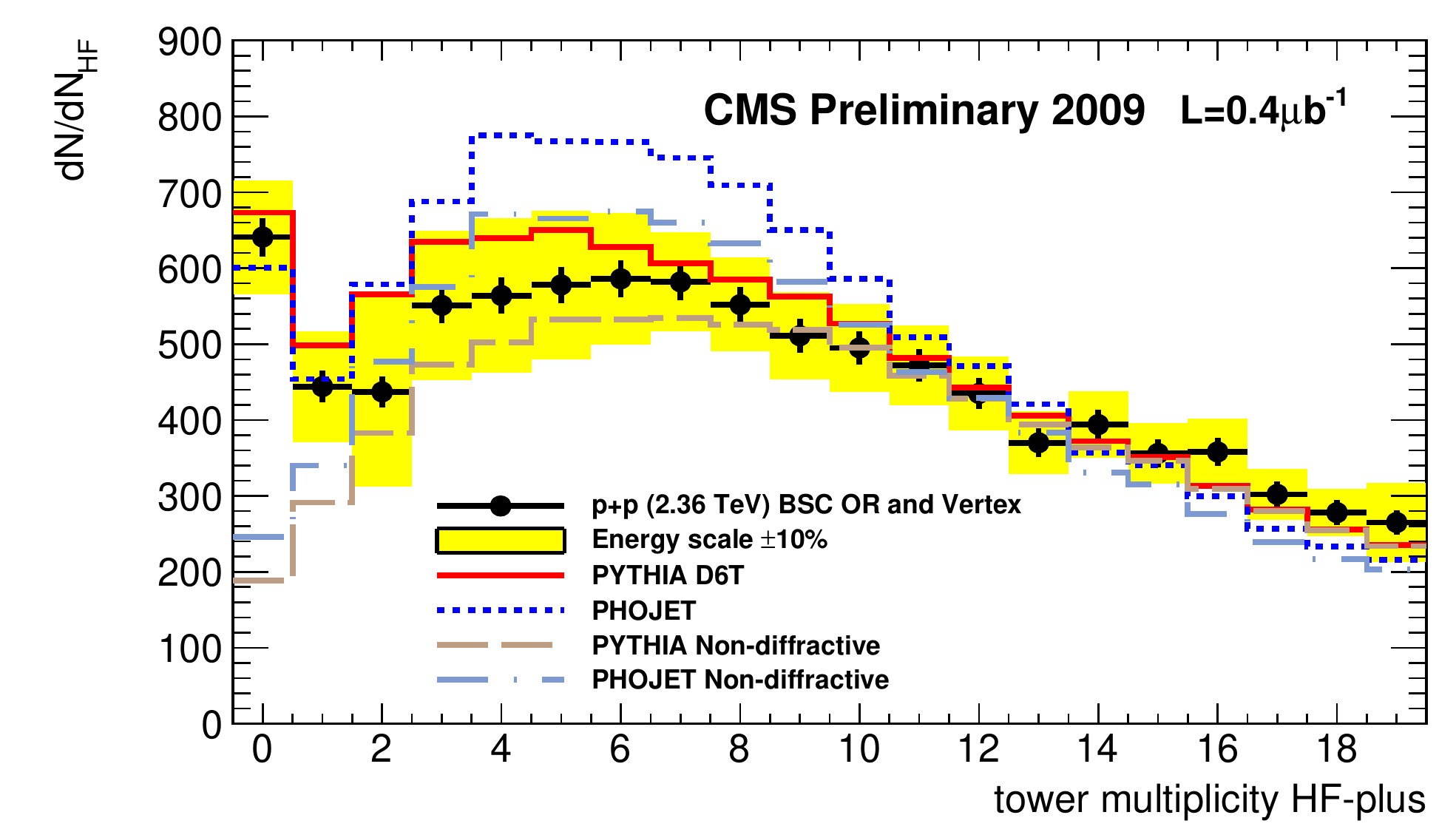} 
   }
   \end{center}
  \caption{\sl Distribution of the uncorrected tower multiplicity in the $HF+$ at $\sqrt{s}=0.9$~TeV~(left) and $2.36$~TeV~(right) . 
               See text for details.}
\end{figure}

Figure~7 shows the distributions of $E - p_{\rm z}$ and $E_{HF-}$ at $\sqrt{s}=0.9$~TeV after applying the cut 
to the energy sum in the HF+ to be below $8$~GeV to enhance diffractive component in the data. 
This additional requirement allows to select mainly SD events with a LRG over HF+ boosting the system $X$ towards the $-z$ direction. 
It can be seen that PHOJET agrees better with the data than PYTHIA after requiring $E_{HF+}<8$~GeV 
particularly giving a better description of the high-mass diffractive systems. 
To complement the study, the selected events were also compared to the Monte Carlo predictions provided 
by three different PYTHIA6 tunes -- D6T, DW and CW, which model multiple parton interactions in a different way. 
It was observed that the tunes give a similar overall description of the data. 
However, due to the present systematic uncertainties it is impossible to discriminate between the tunes.    
\begin{figure}[b!]
  \begin{center}
     \resizebox{0.95\textwidth}{!}{
    \includegraphics[width=1.0\textwidth]{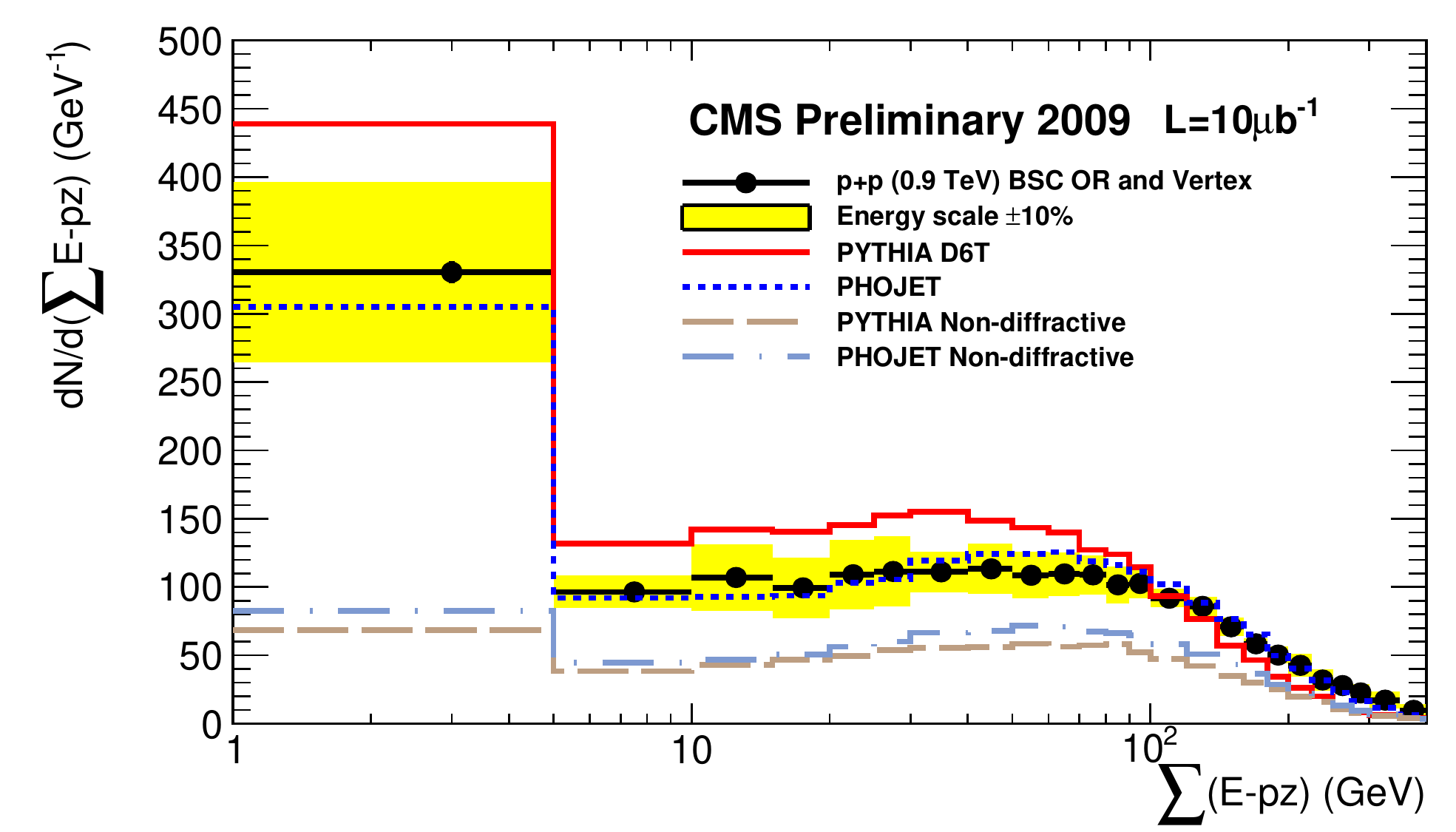}
    \includegraphics[width=1.0\textwidth]{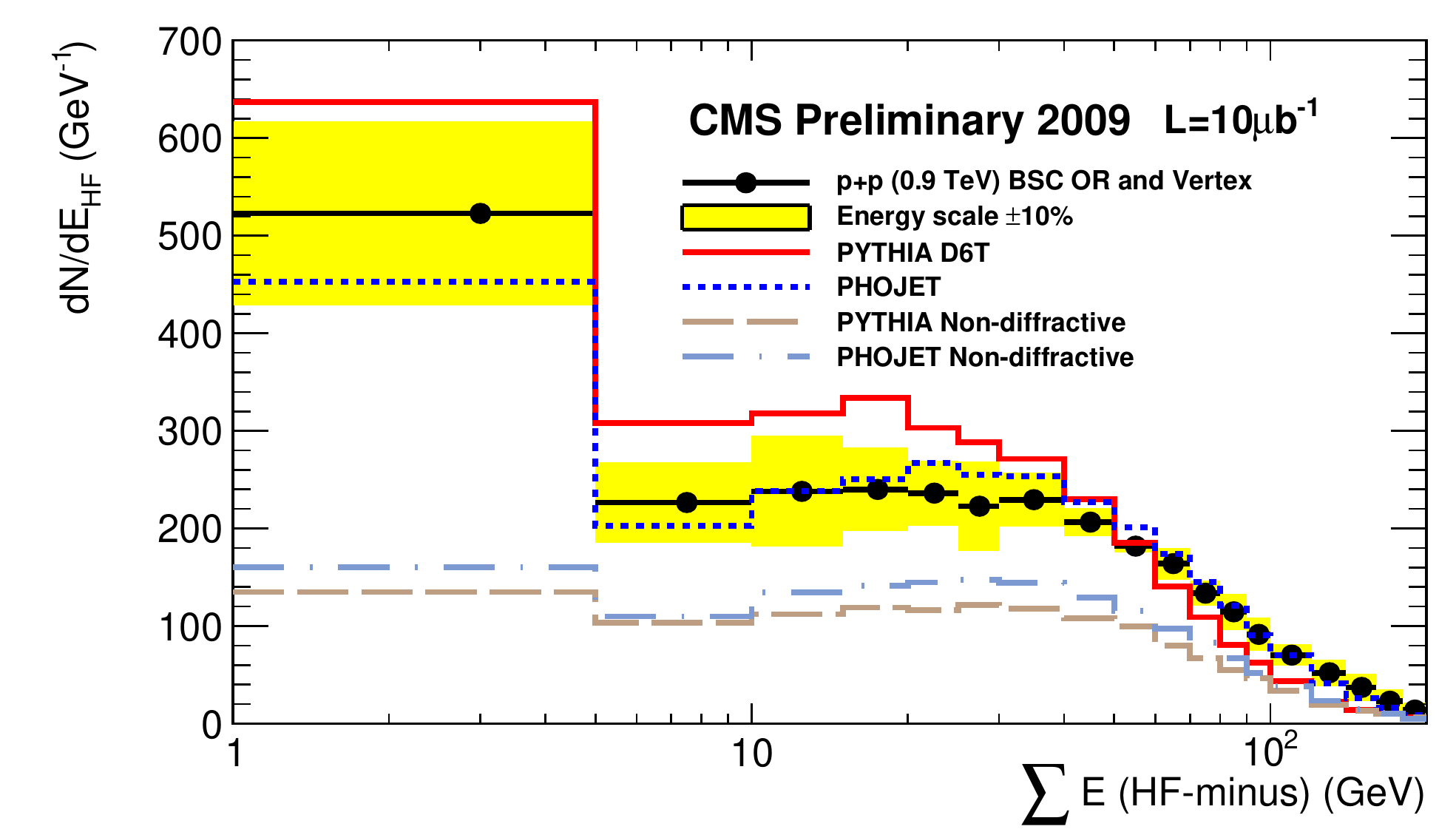} 
   }
   \end{center}
  \caption{\sl Distributions of the uncorrected variables $E-p_{\rm z}$~(left) and $E_{HF-}$~(right) at $\sqrt{s}=0.9$~TeV after applying the additional 
              cut $E_{HF+}<8$~GeV. See text for details.}
\end{figure}
%

\section{Observation of diffractive di-jets}
One of the hard diffractive processes that have been observed by the CMS detector 
throughout 2010 in $pp$ collisions at $\sqrt{s}=7$~TeV is the SD production of di-jets, 
whose diagram is shown in Figure~8. As described in Section~I, this is an interesting process
to study due to its sensitivity to the rapidity gap survival probability and the gluon component of the proton dPDF. 
One of the first candidates of such a process recorded at $\sqrt{s}=7$~TeV~[13] is illustrated in Figure~9. 
The displayed event includes two jets -- one with $p_{\rm T}$ of $43.5$~GeV and $\eta$ of $0.83$ 
and the other with $p_{\rm T}$ of $36.9$~GeV and $\eta$ of $2.55$, which both were reconstructed using the anti-$k_{t}$ algorithm~[14] 
with the radius parameter of $0.5$. In this particular event, no energy deposition in the HF above $4$~GeV is present. 
Moreover, it does not contain any activity in the region $\eta<0$, which is a direct evidence of a LRG that extends over the $-z$ side of the CMS detector. 
This indicates the fact that the proton, which was traveling towards the $+z$ direction, dissociated into the two observed jets,  
whereas the other incoming proton escaped through the beam-pipe towards the $-z$ direction.  
\begin{figure}
\begin{center}
\resizebox{3.0in}{!}{
\rotatebox{0}{
\includegraphics{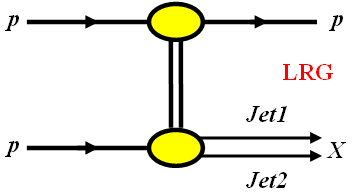}}}
\caption{ \sl Sketch of the SD reaction $pp \rightarrow p X$, where $X$ represents a di-jet system.}
\end{center}
\end{figure}
\begin{figure}
\begin{center}
\resizebox{4.5in}{!}{
\rotatebox{0}{
\includegraphics{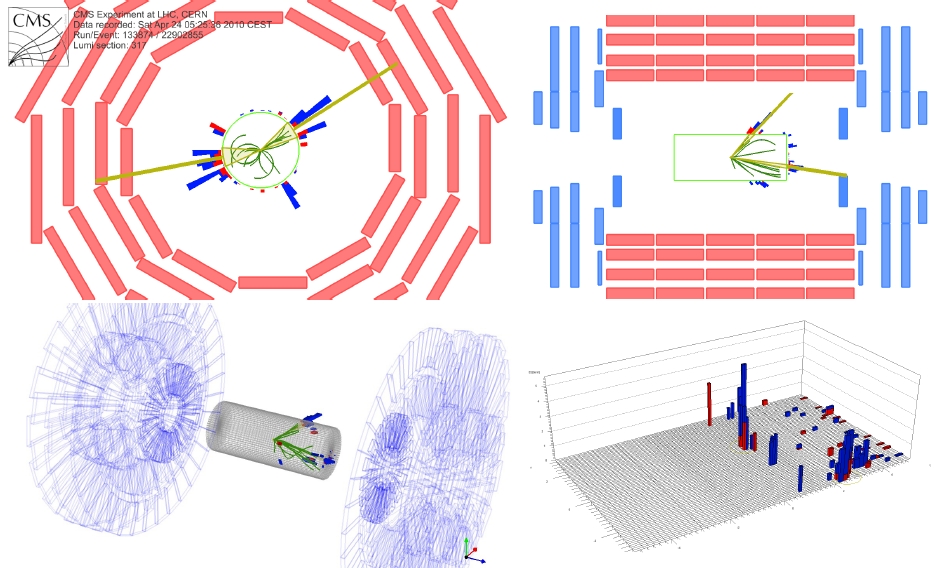}}}
\caption{ \sl Display of an event with diffractive di-jet observed by the CMS detector at $\sqrt{s}=7$~TeV.}
\end{center}
\end{figure}
%

\section{Conclusions}

Inclusive diffraction has been observed by the CMS detector
in minimum bias events collected at $\sqrt{s}=0.9$~TeV and $2.36$~TeV.
At both energies a clear diffractive contribution is evident. 
The data are compared to Monte Carlo predictions provided
by PYTHIA6 and PHOJET generators. It is observed that PHOJET
reproduces the diffractive contribution more accurately than PYTHIA6,
whereas the latter gives a better description of the non-diffractive component of the data.
In addition, first observation of di-jets created in consequence
of the single-diffractive reaction in $pp$ collisions at $\sqrt{s}=7$~TeV is presented.

\section{Acknowledgments}

I would like to express my gratitudes to all people
working in the CMS forward physics community.
Special thanks go to Hannes Jung and Kerstin Borras
for fruitful discussions, suggestions and encouragements.


\end{document}